\documentclass[english,11pt,oneside]{article}

\pdfoutput=1
\usepackage[T1]{fontenc}
\usepackage[latin9]{inputenc}
\usepackage[usenames,dvipsnames]{xcolor}
\usepackage{color}
\usepackage{babel}
\usepackage{setspace}
\usepackage{amstext}
\usepackage{amssymb}
\PassOptionsToPackage{normalem}{ulem}
\usepackage{ulem}
\usepackage[unicode=true,pdfusetitle,
 bookmarks=true,bookmarksnumbered=false,bookmarksopen=false,citecolor=Turquoise,
 breaklinks=false,pdfborder={0 0 1},backref=false,colorlinks=true,
linktoc=page]
 {hyperref}
\usepackage{graphicx}
\usepackage{mathtools}
\usepackage[small]{caption}
\usepackage{changepage}
\usepackage{slashed}
\usepackage{makecell}
\usepackage{multirow}
\definecolor{note_fontcolor}{rgb}{0.80078125, 0.80078125, 0.80078125}
\usepackage[top=3 cm, bottom=3 cm, left=3.1416 cm, right=3.1416 cm]{geometry}

\def\beq{\begin{equation}}
\def\eeq{\end{equation}}
\def\bea{\begin{eqnarray}}
\def\eea{\end{eqnarray}}

\begin{document} 

\baselineskip=17pt


\thispagestyle{empty}
\vspace{20pt}
\font\cmss=cmss10 \font\cmsss=cmss10 at 7pt

\begin{flushright}
UMD-PP-018-01 \\
\end{flushright}

\begin{center}
{\Large \textbf
{Hybrid seesaw leptogenesis and TeV singlets 
}}
\end{center}

\vspace{15pt}

\begin{center}
{\large Kaustubh Agashe$\, ^{a}$, Peizhi Du$\, ^{a}$, Majid Ekhterachian$\, ^{a}$, Chee Sheng Fong$\, ^{b,b*}$,\\ Sungwoo Hong$\, ^{c}$, and~Luca Vecchi$\, ^{d}$ \\
\vspace{15pt}
$^{a}$\textit{Maryland Center for Fundamental Physics,
     Department of Physics,
     University of Maryland,
     College Park, MD 20742, U.~S.~A.} \\
$^{b}$\textit{Instituto de F\'isica, Universidade de S\~ao Paulo, S\~ao Paulo, Brazil} \\
$^{b*}$\textit{Departamento de F\'isica, Pontif\'icia Universidade Cat\'olica do Rio de Janeiro, Rio de Janeiro, Brazil} \\
$^{c}$\textit{Department of Physics, LEPP, Cornell University, Ithaca NY 14853, U.~S.~A.} \\
$^{d}$\textit{Theoretical Particle Physics Laboratory, Institute of Physics, EPFL, Lausanne, Switzerland}
   
\vspace{0.3cm}
      
{\it email addresses}: kagashe@umd.edu, pdu@umd.edu, ekhterachian.majid@gmail.com, 
cheesheng.fong@gmail.com, sungwoo83hong@gmail.com, luca.vecchi@epfl.ch}

\end{center}

\vspace{5pt}

\begin{center}
\textbf{Abstract}
\end{center}
\vspace{5pt}

{\small{
The appealing feature of inverse seesaw models is that the Standard Model (SM) neutrino mass emerges from the exchange of TeV scale singlets with sizable Yukawa couplings, which can be tested at colliders. However, the tiny Majorana mass splitting between TeV singlets, introduced to accommodate small neutrino masses, is left unexplained. Moreover, we argue that these models suffer from a structural limitation that prevents a successful leptogenesis if one insists on having unsuppressed Yukawa couplings and TeV scale singlets. In this work we propose a {\emph{hybrid seesaw}} model, where we replace the mass splitting with a coupling to a high scale seesaw module including a TeV scalar. 
We show that this structure achieves the goal of filling {\em both} the above gaps with couplings of order unity. The necessary structure automatically arises embedding the seesaw mechanism in composite Higgs models, but may also be enforced by new gauge symmetries in a weakly-coupled theory. Our hybrid seesaw models have distinguishing features compared to the standard high scale type-I seesaw and inverse seesaw. Firstly, they have much richer phenomenology. Indeed, they generally predict new TeV scale physics (including scalars) 
potentially accessible at present and future colliders, whereas weakly-coupled versions may also have astrophysical and cosmological signatures due to the presence of a light Nambu-Goldstone boson coupled to neutrinos. Secondly, our scenario features an interesting interplay between high scale and TeV scale physics in leptogenesis and enlarges the range of allowed high scale singlet masses beyond the usual {$\sim 10^9-10^{ 15 }$} GeV, without large hierarchies in the Yukawa couplings nor small mass splitting among the singlets. 
}}

\vfill\eject
\noindent

\section{Introduction}
%
The seesaw mechanism \cite{original} elegantly explains the extreme smallness of the SM neutrino masses. At the same time, the Majorana nature of SM neutrino, 
i.e., the presence of lepton-number violation, raises the highly attractive possibility of baryogenesis via leptogenesis \cite{Fukugita:1986hr,Davidson:2008bu}. 
In this work, we consider a class of seesaw models called inverse seesaw \cite{inverse}. We first emphasize two inadequacies of the standard inverse seesaw scenario and then build an extended framework, which we will term \emph{hybrid seesaw}, to overcome both issues.

\section{Inverse seesaw: $\mu$-problem and leptogenesis}\label{ISS}
%
In the inverse seesaw one introduces a Dirac SM singlet (made up of two Weyl spinors: $\Psi$ and $\Psi^c$) supplemented with an additional {\em tiny} Majorana mass term for one of the chiralities and Yukawa coupling of the {\em other} chirality to the SM Higgs and lepton doublet (denoted by $H$ and $\ell$ respectively):
\bea
-{\cal L} \supset  y \Psi^c H \ell + m_{ \Psi } \Psi \Psi^c + \frac{\mu}{2} \Psi \Psi + {\rm h.c..}
\eea
The generation indices have been suppressed for brevity ($y,m_\Psi,\mu$ are in general matrices). Here $m_{ \Psi }$ is assumed to be in the TeV range, while $\mu\ll m_\Psi$. Integrating out these {\em pseudo}-Dirac singlets generates a small neutrino mass:
\bea
m_{ \nu } \sim \frac{ ( y \; v )^2 }{ m_{ \Psi }^2 } \mu,
\label{mnu_IS}
\eea
where $v = 174$ GeV is the vacuum expectation value (VEV) of $H$.
The crucial point here is that we can obtain the observed size of the SM neutrino mass $m_\nu \sim 0.05$ eV with 
unsuppressed Yukawa couplings $y =O(0.01-1)$ and $m_\Psi\sim1$ TeV, provided $\mu =O( \textrm{10 keV} - \textrm{eV})$.

An attractive feature of this scenario is that the singlets are potentially accessible at colliders because of their unsuppressed Yukawa coupling \cite{Mohapatra:2016twe}.\footnote{The singlets may also be charged under new gauge symmetries broken at the TeV scale, giving additional production channels.} However, this set-up also has two drawbacks. Firstly, if the new physics resides at the TeV scale, there is a priori no reason to expect $\mu$ in the keV range or below. Although a small Majorana mass term $\mu$ is technically natural (since a symmetry, namely lepton number, is restored by its vanishing), the required value appears as unexpected within this picture: additional ingredients are needed. Secondly, as we will argue next, it appears difficult to achieve successful leptogenesis 
in this framework.

To study leptogenesis, we first calculate the CP asymmetry from decays of heavy singlet, 
\bea
\epsilon_ \Psi  \equiv\frac{|\Gamma_\Psi-\overline\Gamma_\Psi|}{\Gamma_\Psi+\overline\Gamma_ \Psi },
\eea
where $\Gamma_{ \Psi }(\overline\Gamma_\Psi)$ is the decay width of $\Psi$ into $\ell H (\ell^* H^\ast)$. Assuming anarchic $\mu$-terms and singlet masses and no hierarchies in Yukawa couplings, we have: 
\bea\label{asymIS}
\epsilon_ \Psi   \sim  
\frac{ \mu }{ m_{ \Psi } } \frac{ \mu }{ \Gamma_{ \Psi } },
\eea
where the first factor may be interpreted as arising from the CP phase in Yukawa couplings, whereas the second comes from the on-shell propagator  due to the near-degeneracy of the pseudo-Dirac pair $\Psi$, $\Psi^c$ when calculating one-loop self-energy corrected decay width. The two powers of $\mu$ in eq.~\eqref{asymIS} can be understood in generality using the Nanopoulos-Weinberg theorem \cite{Nanopoulos:1979gx}, that states that we need to go to at least second order in the lepton-number breaking parameter (namely, the $\mu$-term in this model) in order to generate an asymmetry. 
This result was first obtained in the first reference in \cite{Deppisch:2010fr} and 
was backed up by a detailed analysis \cite{Deppisch:2010fr}. Crucially, it depends on the regulator used for the almost on-shell $\Psi$ propagator in the self-energy diagram.

To determine the present-day asymmetry we should combine the above result with the effective washout factor from the {\em inverse} decay of SM leptons and Higgs into the singlets. This latter
quantity was first estimated in \cite{Blanchet:2009kk}:
\bea
K^{  \rm eff }_{ \Psi } \sim \frac{\Gamma_\Psi}{H_\Psi} \frac{\mu^2}{\Gamma^2_\Psi},
\label{washout_IS} 
\eea
where $K_{ \Psi } \sim \Gamma_{ \Psi } / H_{ \Psi }$ is the ``usual'' washout factor \cite{Davidson:2008bu} and $H_{ \Psi }$ is the Hubble parameter
at $T=m_\Psi$, i.e., $H_\Psi\sim \sqrt{ g_{ \ast } } m_{ \Psi }^2 / M_{ \rm Pl }$, with $g_{ \ast }$ being the number of relativistic degrees of freedom at that temperature and $M_{\rm Pl}$ the Planck mass. The quadratic suppression in $\mu$ comes from the fact that the rate for lepton-number-violating processes,
e.g. $\ell H\leftrightarrow(\ell H)^*$, should vanish
in the lepton-number conserving limit.\footnote{Throughout the paper we take $\mu \ll \Gamma_\Psi$,
as is expected given that Yukawa couplings are unsuppressed.} If for definiteness we focus on the strong washout region, in which $K^{  \rm eff }_{ \Psi } \gg 1$, the net lepton asymmetry can be obtained as\footnote{The superscript $\Psi$ is to remind the reader that the asymmetry originates from decays of $\Psi$. To be precise one should refer to the $B-L$ charge. However for simplicity we will work with a lepton asymmetry.} 
\bea
Y^{ \Psi }_{ \Delta \ell } \sim10^{-3} \frac{ \epsilon_{ \Psi } }{ K^{  \rm eff }_{ \Psi} } ,
\label{general_Y}
\eea
where $Y_X\equiv n_X/s $ ($Y_{\Delta X}\equiv (n_X-n_{X^*})/s $) with $n_X$ being number density of the corresponding species and $s$ being total entropy {density} of the Universe. The numerical factor $\sim10^{-3}$ in eq.~(\ref{general_Y}) comes from relativistic number density of $\Psi$ normalized to $s$. 

Putting everything together, and assuming strong washout for simplicity, we find
\bea
Y^{ \Psi }_{ \Delta \ell } \sim10^{-3}  \frac{ \sqrt{ g_{ \ast } } m_{ \Psi } }{ M_{ \rm Pl } }\sim 10^{-18}\left(\frac{g_{\ast}}{100}\right)^{\frac{1}{2}}\left(\frac{m_\Psi}{\textrm{TeV}}\right).
\label{net_IS}
\eea
The final lepton asymmetry in eq.~(\ref{net_IS}) is {\em in}dependent of the size of the Yukawa couplings. Furthermore, given that $Y_{\Delta B}\sim Y^{ \Psi }_{ \Delta \ell }$ after electroweak sphaleron processes are taken into account, we see that eq.~(\ref{net_IS}) predicts too small baryon asymmetry to account for the observed one ($Y^{\rm obs}_{\Delta B} \sim 10^{-10}$) for singlet masses in the TeV ballpark. In order to reach this conclusion, it is important to include the effect of washout: considerations based solely on $\epsilon_\Psi$ could suggest that larger $\mu$ than benchmark value shown below eq.~(\ref{mnu_IS}) might suffice. For example, taking $\mu \sim 10$ MeV (and compensating this increase by reducing the size of $y$ to $y \sim$ a few $10^{-3}$ to keep $m_\nu$ fixed) gives rise to $\epsilon_\Psi \sim 10^{-7}$. While the difficulty in getting required size of CP violation was pointed out in \cite{Deppisch:2010fr}, to our knowledge, the parametric form of eq.~(\ref{net_IS}) including washout effect has never been presented before.

A small baryon asymmetry is a very generic implication of TeV scale inverse seesaw. We will show in a companion paper \cite{long} that even allowing a departure from the above generic conditions, for example allowing a degeneracy among different generations of singlets, as well as considering the weak washout regime, the inverse seesaw scenario can at most barely reach the required asymmetry. Introducing other small sources of lepton-number violation as in the linear seesaw model \cite{Malinsky:2005bi} does not change this conclusion \cite{long}. Similar conclusions are obtained in  the numerical analysis of ref.~\cite{Dolan:2018qpy}.

\section{A hybrid seesaw model}

We now construct an extension of the original inverse seesaw model that features a high-scale module. We will see that, if the interactions between the low and high scale modules are properly chosen, the resulting scenario can simultaneously address {\em both} the smallness of neutrino masses and leptogenesis.

Our model is the following:
\bea\label{model}
- {\cal L} \supset y \Psi^c H \ell + \kappa \Phi_{ \kappa } \Psi \Psi^c + \lambda \Phi_{ \lambda } \Psi N + \frac{M_N}{2} N N+{\rm h.c.}.
\label{toy-model}
\eea
Here $N$ is a {\em super}-heavy singlet with mass $M_N\gg$ TeV, whereas $\Psi,\Psi^c,\Phi_{\lambda,\kappa}$ acquire masses 
(and VEV's) of the order of TeV. Following the philosophy of inverse seesaw, we work with unsuppressed Yukawa couplings $y,\lambda,\kappa$.
Furthermore, we will assume anarchical Yukawa couplings 
such that different generations are comparable, complex phases are of order unity, and the masses of $N$ are not hierarchical
nor quasi-degenerate (i.e., $M_{N_1} \lesssim M_{N_2}$)\footnote{A realistic neutrino mass matrix and leptogenesis both require at least two generations of $N$.} so that we can simply suppress the generation indices in our expressions.\footnote{Having said this, most of our results will be valid even in the case of hierarchical masses and couplings. In the companion paper \cite{long}, we will study the latter situation in more detail.}

Importantly, our model is not just a random merger of standard type-I and inverse seesaw. Indeed, a few non-trivial conditions have to be satisfied in order to obtain the scenario shown in eq.~(\ref{model}). Firstly, while it is the large Majorana mass $M_N$ that furnishes the  number-breaking necessary to generate neutrino masses and leptogenesis, $N$ does not couple directly to the SM. In order to realize the inverse seesaw mechanism, it is crucial that the number-breaking is communicated to the SM via lighter degrees of freedom ($\Psi,\Psi^c,\Phi_{\lambda,\kappa}$), which thus act as {\emph{mediators}}. We will see below that $\lambda\langle\Phi_\lambda\rangle, \kappa\langle\Phi_\kappa\rangle\sim$ TeV $\ll M_N$ allows us to obtain realistic neutrino masses. Secondly, successful leptogenesis requires a {\emph{dynamical}} $\Phi_\lambda$. In addition, $\Phi_{\lambda,\kappa}$ can be either both real or complex. We will see below that, in the latter case, their potential should either break all the global symmetries of eq.~(\ref{model}), or respect them all. If a linear combination is preserved the primordial asymmetry may be washed out~\cite{long}. Remarkably, we will see that these conditions are {\emph{automatically}} realized once the non-generic coupling structure of eq.~(\ref{model}) is enforced by the UV dynamics that we consider below.

The characteristic structure in eq.~(\ref{model}) may be enforced introducing a $U(1)_{B-L}\times U(1)_X$ gauge symmetry \cite{long}. In the simplest such realization, $N$ appears in two generations, which coincides with the minimal number of generations necessary to obtain a realistic neutrino mass matrix. Furthermore, the very same structure is naturally realized embedding the standard type I seesaw in Composite Higgs (CH) scenarios, dual to warped extra dimensions \cite{Agashe:2015izu}. In such a composite seesaw, the peculiar couplings of eq.~(\ref{model}) arise because the fields $\Psi$, $\Psi^c, H, \Phi_{ \kappa, \lambda }$ are identified as resonances of a strongly-coupled sector (the role of $\Phi_{ \kappa, \lambda }$ is played by the dilaton). On the other hand, $N$ and $\ell$ are states in an elementary sector external to the strong dynamics. They couple to the strong sector via the {\emph{mixing}} with composite fermionic resonances. For example, the combination $\Psi\Phi_\lambda$ plays the role of the composite fermionic operator mixing with $N$; the fermionic resonance inducing a coupling between $\ell$ and the Higgs ($y$-term) is not explicitly shown above because it does not play any key role in our analysis. Importantly, this picture forbids direct couplings between $N$ and the SM leptons (cf. in standard type-I seesaw).

A model similar to eq.~(\ref{model}) was previously considered in \cite{Aoki:2015owa}. 
The crucial difference is that those authors considered $M_{N} = O ({\rm TeV})$ and $\lambda\langle\Phi_\lambda\rangle=O (10)$ MeV. On the other hand, in our paper we take $M_N\gg$ TeV and $\lambda\langle \Phi_\lambda \rangle =O({\rm TeV})$. Our choice not only allows a natural explanation for the smallness of $\mu \sim $ keV, and is {\emph{necessary}} to generate a realistic baryon asymmetry.

\subsection{Solution to the $\mu$-problem}

Integrating out the super-heavy $N$, and assuming a non-zero $\langle \Phi_{ \lambda } \rangle\sim$ TeV, a super-small effective $\mu$-term for the TeV mass singlets is generated due to a {\em high}-scale seesaw
structure:
\bea\label{mu_eff}
\mu \sim \frac{ \left( \lambda \langle \Phi_{ \lambda } \rangle \right)^2 }{ M_N }.
\eea
In the same spirit as the standard type-I seesaw scenario, the existence of a new heavy scale $M_N\gg$ TeV allows us to transmute the TeV scale into a small dimensionless coupling TeV$/M_N\ll1$. The smallness of neutrino masses appears as a natural consequence of the two scales of our model, thus evading the first concern of the original inverse seesaw model. A Majorana mass splitting of the right order of magnitude [$\mu=O({\rm eV - 10~keV})$] is here obtained with $\lambda\langle\Phi_{ \lambda } \rangle = O ( \hbox{TeV} )$ 
and $M_N$ at a scale which is intermediate between TeV and Planck.

Despite the obvious analogy with the standard type-I seesaw, the role of the super-heavy singlet is not exactly the same. In our model, there is no direct contribution to the neutrino mass from integrating out 
$N$. The small lepton number violation is encapsulated at low scales by a small parameter $\mu/m_\Psi\ll 1$ and ``communicated'' to the SM via new TeV scale fermions. This represents the {\em hybrid} seesaw structure, i.e., a combination of the high-scale and the TeV inverse seesaw, as is manifest in the expression of the SM neutrino mass in terms of the fundamental parameters [plugging eq.~(\ref{mu_eff}) into eq.~(\ref{mnu_IS})]:
\bea
m_{ \nu } \sim \Big[ \frac{ ( y v )^2 }{ M_N } \Big] \left( \frac{ \lambda \langle \Phi_{ \lambda } \rangle }{  \kappa \langle \Phi_{  \kappa } \rangle }  \right)^2.
\label{mnu_toy}
\eea
The first factor in eq.~(\ref{mnu_toy}) is the usual high-scale seesaw expression, whereas the second is a ``modulation'' due to the TeV-scale physics acting as a link to the SM. In a warped extra-dimensional picture of our model, where $\Psi,\Psi^c$ are the Kaluza-Klein excitations of a 5D field with UV boundary value $N$, the latter factor is controlled by the wavefunction of the bulk singlet~\cite{Agashe:2015izu}. This is itself a dual description of a renormalization group effect 
in 4D~\cite{Agashe:2015izu}.
Note that this effective ``modulation'' factor in warped/composite model can be naturally (much) smaller or larger than $O(1)$, i.e., 
with{\em out} invoking any hierarchies in the {\em fundamental} (whether 5D or 4D) parameters.

\subsection{A two-step leptogenesis}
%

While the VEV of the new scalars are enough to generate neutrino masses, a realistic model for leptogenesis requires that $\Phi_{\lambda}$ be a dynamical field.
This key ingredient opens the possibility to the following decays:
\bea
N \rightarrow \Psi \Phi_{ \lambda }, (\Psi \Phi_{ \lambda })^*,
\label{Ndecay}
\eea
that can potentially create an asymmetry at temperatures of the order of $M_N$. 

\begin{table}[t]
\begin{center}
\begin{tabular}{c|cc} 
\rule{0pt}{1.2em}%
 & $U(1)_{B-L}$ & $U(1)_{\lambda}$ \\
 \hline
  $\ell$ &  $-1$ & $1$   \\
 $\Psi$ &  $0$ & $1$   \\
 $\Psi^c$ &  $+1$ & $-1$   \\
  $N$ &  $0$ & $0$ \\
  $\Phi_\kappa$ & $-1$ & $0$ \\
 $\Phi_\lambda$ & $0$ & $-1$ 
\end{tabular}
\caption{\small{Charge assignments under the two global symmetries of eq.~(\ref{model}). In our UV completion based on \emph{gauge} symmetry $U(1)_{B-L}\times U(1)_X$ \cite{long}, these arise as accidental. None of the two global symmetries is expected in the case of a UV completion via warped extra dimensions, where $\Phi_{\kappa,\lambda}$ are real and identified with the dilaton.\label{tab}}}
\end{center}
\end{table}

To obtain a successful model it is necessary that the asymmetry does not get washed-out by later reactions. To identify the key washout processes one should first understand the symmetries of the model. Now, if $\Phi_{\lambda,\kappa}$ are complex the interactions of eq.~(\ref{model}) enjoy two global symmetries $U(1)_{B-L} \times U(1)_\lambda$ (see table~\ref{tab}). We find that successful realizations of leptogenesis are naturally achieved if the two global symmetries shown in table~\ref{tab} are also satisfied by the scalar potential, and therefore are symmetries of full Lagrangian, or if they are completely broken, for example because the potential is generic or the scalars are real. Whether or not one can obtain a realistic baryon asymmetry in models in which the potential preserves a linear combination of the symmetries in table~\ref{tab} is more model-dependent and will not be analyzed here.

In the following we will therefore discuss leptogenesis in fully-symmetric models, where $U(1)_{B-L}\times U(1)_\lambda$ are symmetries of the full theory, as well as in non-symmetric models, where the symmetry is completely broken. Interestingly, the former case is automatically realized in the \emph{gauge} $U(1)_{B-L} \times U (1)_X$ model \cite{long} introduced to motivate the structure of eq.~(\ref{model}), since the assignment of gauge charges therein forbids couplings of the form $\Phi^n_{\kappa} \Phi^m_{\lambda}$. In that model, while the $U(1)_{B-L}$ is actually gauged, the $U(1)_\lambda$ arises as an accidental global symmetry and is not to be identified with the gauge $U(1)_X$. This is the class of {\emph{fully-symmetric}} models we will discuss in the following. On the other hand, in the other justification for eq.~(\ref{model}) that we provided, i.e., composite Higgs scenarios, one automatically falls in the class of {\emph{non-symmetric}} models, since $\Phi_{\kappa,\lambda}$ are replaced by a single real scalar, i.e., the dilaton.

\paragraph{Fully-symmetric models}

We begin with a discussion of fully-symmetric models, in which the $U(1)_{B-L}$ is gauged (this also implies an exact global $B-L$) while $U(1)_\lambda$ is 
a global symmetry. We first consider the regime $T\gg$ TeV, where the new scalars $\Phi_{ \kappa, \lambda }$ are assumed to have vanishing VEVs. This is a generic possibility because thermal effects usually stabilize the field origin. 
In such a regime, the yields $Y^N_{\Delta i}$ of the different species $i=\ell,\Psi,\Psi^c,\Phi_\lambda,\Phi_\kappa$ satisfy 
\bea
Y^N_{\Delta\ell}+Y^N_{\Delta\Psi}-Y^N_{\Delta\Psi^c}&=&Y^N_{\Delta\Phi_\lambda}, \label{charges} \\
Y^N_{\Delta\ell}-Y^N_{\Delta\Psi^c}+Y^N_{\Delta\Phi_\kappa}&=&0. \nonumber
\eea
In addition to the two global symmetries, there is an {\emph{approximate}} lepton number under which only $\ell,\Psi,\Psi^c$ and $N$ are charged, which is violated by the couplings to $N$. The decay of $N$ [eq.~(\ref{Ndecay})] generates an asymmetry in the latter quantity,
\bea\label{appcharges}
Y^N_{\Delta\ell}+Y^N_{\Delta\Psi}-Y^N_{\Delta\Psi^c}\neq0,
\eea
while maintaining a vanishing value for the two exactly conserved charges [see eq.~(\ref{charges})]. This illustrates that, while eq.~(\ref{charges}) specifies two conservation laws, the left and right hand sides of the first equation are not separately conserved.

To retain a non-vanishing $Y^N_{\Delta\ell}$ at late times, two conditions have to be guaranteed. First, all processes that deplete $Y^N_{\Delta\ell}$ must be suppressed. Second, all reactions removing $Y^N_{\Delta\Phi_\lambda}$ must be inefficient as well. Indeed, if $Y^N_{\Delta\Phi_\lambda}\to0$ happens still in the symmetric phase, before $\Phi_\lambda$ acquires a VEV, comparing eq.~(\ref{appcharges}) and the first relation in eq.~(\ref{charges}) one finds that the $\ell$ asymmetry gets completely depleted after $\Psi,\Psi^c$ decay (as long as this occurs before electroweak sphalerons shut off). Let us see what are the conditions necessary to avoid such depletion.

First, note that in fully-symmetric models, and assuming $M_N$ is much heavier than the other particles, there are no processes that can remove $Y^N_{\Delta\Phi_\lambda}$. This is a consequence of the fact that eq.~(\ref{charges}) forces such processes to involve an odd number of fermions and is therefore forbidden when combined with Lorentz invariance. In fact, $\Phi_{\lambda}$ decays only after it acquires a VEV. To see that it is exactly stable when $U(1)_\lambda$ is preserved, note that all decay products allowed by the global symmetry are forced to contain one lepton charge (see table~\ref{tab}) plus, by Lorentz invariance, an odd number of fermions that are total singlets under all internal symmetries. The only possible option in our model is $N$, but this is kinematically forbidden under our working hypothesis $M_N\gg$ TeV. This automatically prevents a very dangerous type of $Y^N_{\Delta\Phi_\lambda}$ depletion. The only reactions that can washout $Y^N_{\Delta\ell}$ and $Y^N_{\Delta\Phi_\lambda}$ are therefore scattering processes, which we analyze next.

In our scenarios the dominant number-changing interactions at $T\gg$ TeV arise from the usual inverse decay processes $\Phi_{ \lambda } \Psi, ({ \Psi } { \Phi }_{ \lambda })^*\to N$ at $T\sim M_N$. This is 
parametrized by the washout factor \cite{Davidson:2008bu}~\footnote{In the case genesis occurs at $M_N\gtrsim 10^{15}$ GeV, the fields $\Psi, \Phi_\lambda, N$ may be kept in thermal equilibrium by a sizable $\lambda$, but the SM is typically decoupled. In such case, we assume inflaton only populates the singlet sector, resulting in $g_*\sim10$.}
\bea
K_N \sim \frac{ \Gamma_N }{ H_N } 
\simeq 2 \left(\frac{\lambda}{0.5}\right)^2
\left(\frac{10}{g_\ast}\right)^{1/2}
\left(\frac{10^{16}\,{\rm GeV}}{M_N}\right),
\label{KN}
\eea
where we used $\Gamma_N \sim M_N \lambda^2 / \left( 16 \pi \right)$ for the decay width of $N$ and $H_N \sim \sqrt{ g_{ \ast } } M_N^2 / M_{ \rm Pl }$ is the Hubble parameter at $T\sim M_N$. Off-shell scatterings $\Phi_{ \lambda } \Psi { \leftrightarrow } ({ \Psi } { \Phi }_{ \lambda })^*$, still mediated by the coupling $\lambda$, can deplete $Y^N_{\Delta\ell}$ (and $Y^N_{\Delta\Psi}$) at even lower temperatures and should be suppressed. For these to be ineffective we require $\Gamma_{ \rm scattering } \sim \frac{\lambda^4 T^3 }{ 16\pi^3 M_N^2 } < H(T)$,\footnote{We use $n(T)\sim T^3/ \pi^2$ for the number density of relativistic particles in thermal equilibrium.} a condition that can be conservatively written, by setting $T \sim M_N$, as
\bea
\frac{ \lambda^4 }{ 16\pi^3 } < \sqrt{ g_{ \ast } }\frac{ M_N }{ M_{ \rm Pl } }
\implies K_N <  9\left(\frac{10}{g_\ast}\right)^{1/4}\left(\frac{10^{16}\,{\rm GeV}}{M_N}\right)^{1/2},
\label{scatt_wash}
\eea
where in the second part of the above equation we have used eq.~(\ref{KN}).

Having identified the condition eq.~(\ref{scatt_wash}) to avoid washout at high temperatures, we should now consider what happens below the critical temperature $T_c$ at which the scalars acquire VEV's. For simplicity, we assume the two VEV's are comparable and $\kappa = O(1)$, so that $m_{ \Psi }$ is also of roughly similar value, and that the phase transition is smooth, so that no large entropy production occurs. In these fully-symmetric models, the phase transition implies the existence of two massless Nambu-Goldstone bosons (NGBs): the phase of $\Phi_\kappa$ is eaten by the $U(1)_{B-L}$ vector, whereas the one of $\Phi_\lambda$ is physical. The physical $\Phi_\kappa$ component can decay into SM particles via the TeV singlet fermions or the $B-L$ vector. On the other hand, writing $\Phi_\lambda=(\langle\Phi_\lambda\rangle+\phi_\lambda)e^{i\pi_\lambda/\langle\Phi_\lambda\rangle}$, one finds that $\phi_\lambda$ decays into a $\pi_\lambda$ pair promptly. The phenomenology of $\pi_\lambda$ will be discussed in section~\ref{sec:pheno}. 

At these temperatures potentially relevant number-violating interactions turn on. Because the two global symmetries are now broken, the conservation conditions in eq.~(\ref{charges}) no longer hold. There is only an {\emph{approximate}} global symmetry, i.e., the generalized lepton number in eq.~(\ref{appcharges}). The dangerous washout processes are therefore those that directly affect $Y^N_{\Delta\Psi,\Delta\Psi^c,\Delta\ell}$. One example is the operator $\sim \lambda^2 \langle \Phi_{ \lambda } \rangle { \Phi }_{ \lambda } \Psi^2 / M_N$, obtained by integrating out $N$ and setting {\em one} $\Phi_{ \lambda }$ to its VEV. It is simple to show that processes involving dynamical $\Phi_\lambda$ are always out-of-equilibrium as long as eq.~(\ref{scatt_wash}) holds and $\lambda\langle\Phi_\lambda\rangle\ll M_N$. This ensures that the inverse decays $\Psi\Psi\to\Phi_\lambda$, which violate the approximate lepton number in eq.~(\ref{appcharges}), are out of equilibrium. As a consequence, also the scattering $\Psi\Psi\leftrightarrow(\Psi\Psi)^*$ mediated by {\em off}-shell $\Phi_{ \lambda }$ can safely be ignored. 

The relevant washout processes to consider at $T\lesssim T_c$ are $\ell, \Psi,\Psi^c$-changing interactions from setting {\emph{all}} $\Phi_\lambda$ to its VEV, and are therefore controlled by the effective $\mu$-term [eq.~(\ref{mu_eff})]. The dominant ones are the resonant reactions 
$\ell H\leftrightarrow(\ell H)^*$ mediated by on-shell $\Psi,\Psi^c$,\footnote{This is equivalent to inverse decays process of 
$\Psi,\Psi^c \to \ell H, (\ell H)^*$.}
as it is for standard inverse seesaw scenarios discussed in section~\ref{ISS}.

We thus see that the model presents a two-step washout. At high temperatures the main effect is peaked at $T\sim M_N$ from inverse decay of $N$. No additional washout effects during intermediate temperatures are possible if eq.~(\ref{scatt_wash}) holds. Then, new washout processes emerge at $T\sim \lambda \langle \Phi_{ \lambda } \rangle,\kappa \langle \Phi_{ \kappa } \rangle$, and are controlled by the same parameter as in eq.~(\ref{washout_IS}): 
\bea\label{IR}
K^{\rm eff }_{ \Psi } \sim \frac{ 16 \pi }{ y^6 } 
\frac{ M_{ \rm Pl } \; m_{ \nu }^2 \; m_{ \Psi } }
{ \sqrt{ g_{ \ast } } \; v^4 },
\label{IR_washout}
\eea
where we used $\Gamma_\Psi \sim y^2 m_\Psi / (16 \pi)$ and eq.~(\ref{mnu_IS}) to replace $\mu$ in terms of the more physical quantities $m_\nu$, $v$ and $m_\Psi$.
However, note that $K^{\rm eff }_{ \Psi }$ enters the final asymmetry in a completely different way compared to what shown in eq.~(\ref{general_Y}). In the present case, the IR washout $\ell H\leftrightarrow(\ell H)^*$ becomes effective at $T \sim m_\Psi$ and induces an {\emph{exponential}} suppression of the {\em primordial} asymmetry, see eq.~(\ref{final}) below. 
For $T \ll m_{ \Psi } $, all number-changing effects are negligible, and the right-handed side of eq.~\eqref{appcharges} will become constant.
After $\Psi,\Psi^c$ decay into the SM particles, i.e., $Y^N_{\Delta \Psi} = Y^N_{\Delta \Psi^c} = 0$, the lepton asymmetry will fully reside in $Y^N_{\Delta \ell}$.

\paragraph{Non-symmetric models} 

Scenarios with a completely generic potential, or with real $\Phi_{\lambda,\kappa}$, have no exact global symmetries, and no NGBs. In this case last two relations in eq.~(\ref{charges}) do not hold and the decays of $N$ generate directly $Y^N_{\Delta\Psi,\Delta\Psi^c}$, and ultimately $Y^N_{\Delta\ell}$ via reactions controlled by $y,\kappa$, as in eq.~(\ref{appcharges}). Because there is no global charge associated to $\Phi_{\lambda}$, the only dangerous washout processes are those changing the $\ell,\Psi,\Psi^c$-numbers. As discussed above, these are parametrized by the UV and IR washout parameters in eq.~(\ref{KN}) and eq.~(\ref{IR}), respectively. All other effects are negligible as long as eq.~(\ref{scatt_wash}) is satisfied. In the non-symmetric models the net lepton asymmetry is expected to be the same as for the fully symmetric scenarios, up to factors of order unity~\cite{long}.

\subsubsection{Present-day asymmetry}

For both fully symmetric and non-symmetric models, the picture that emerges is qualitatively as follows. At around $T\sim M_N$ an asymmetry is generated via eq.~(\ref{Ndecay}). We can make use of the standard estimate \cite{Davidson:2008bu}:
\bea\label{epsN}
\epsilon_N  \sim \frac{\lambda^2 }{8\pi},
\eea
arising from interference of tree and one-loop diagrams. In (\ref{epsN}),
we have assumed 
flavor/generational ``anarchy'', as mentioned below eq.~(\ref{toy-model}). At this stage the dominant washout effects are parametrized by eq.~(\ref{KN}). The net asymmetry can be written as \cite{Davidson:2008bu}
\bea
Y_{  \Delta \Psi }^{ N } \sim 10^{-3}\times \left\{
\begin{matrix}
 { \epsilon_N } / K_N  ~~~~~~~~~{\rm for}~~K_N \gg 1\\
 { \epsilon_N } K_N^2 ~~~~~~~~~~{\rm for}~~K_N \ll 1
\end{matrix}
\right. ,
\label{net_UV} 
\eea
depending on whether the UV washout is strong ($K_N\gg 1$) or weak ($K_N \ll 1$). In deriving eq.~(\ref{net_UV}) the initial abundance for singlet $N$ was set to zero and its production is controlled \emph{solely} by $\lambda$, whereas thermal number densities for $\Psi$ and $\Phi_\lambda$ during the genesis are assumed (for both strong and weak UV washout).\footnote{The states $\Psi,\Phi_{\lambda}$ might be thermalized either by the couplings to the SM or directly to the inflaton sector. These possibilities will be investigated in \cite{long}.}

As the universe cools down to $T\ll M_N$, the washout due to UV inverse decay becomes exponentially suppressed. Under the hypothesis shown in eq.~(\ref{scatt_wash}), all other number-changing processes are switched off. The assumption of sizable couplings in eq.~(\ref{model}) ensures that the primordial asymmetry is shared among all particle species, thus resulting in non-vanishing yields $Y_{\Delta i}^{ N }\neq0$ of comparable magnitude.

The asymmetries remain approximately constant down to temperatures of order $T\sim$ TeV, when $\Phi_{\lambda,\kappa}$ acquire a VEV and a $\mu$-term is generated. At this point resonant exchange of $\Psi,\Psi^c$ induces an IR washout [eq.~(\ref{IR})] whose effect (as anticipated earlier) is 
to suppress exponentially the lepton asymmetry generated in the UV. 
There would be additional generation of lepton asymmetry from $\Psi$,$\Psi^c$ decay. However, as discussed in section~\ref{ISS}, such asymmetry is too small and has no impact on the final baryon asymmetry and therefore its contribution is neglected. 
Combining the above UV and IR effects gives us an estimate for the present-day baryon asymmetry:  
\bea\label{final}
Y_{ \Delta B } & \sim & Y_{  \Delta \Psi }^N \exp \left( - K^{\rm eff }_{ \Psi } \right).
\label{final-asym}
\eea

Much like in the generation of SM neutrino mass, we see from eq.~(\ref{Ndecay}) 
that it is the TeV-mass singlet sector which carries information of high-scale number-breaking to the SM 
sector, resulting in eq.~(\ref{final-asym}).
This fact ultimately opens the way to a much richer spectrum of options compared to the standard leptogenesis. For example, the magnitudes of the Yukawa couplings $y,\kappa$ 
-- which do enter in the neutrino mass in eq.~(\ref{mnu_toy}) --
have nothing to do with the generation of the primordial asymmetry, but govern the washout in the IR through eq.~\eqref{IR}. Furthermore, there exists potentially four very different realizations of 
this leptogenesis framework (since we have either strong or weak washout for each of UV and IR components), 
which can impact the final asymmetry in a considerable way. 
We can choose the parameters in such a way that the UV asymmetry is already of roughly the right size and require a weak washout in the IR or, alternatively, start from a large UV asymmetry that is later diluted appropriately by strong washout in the IR. 
We will show these features in the next section.

\section{Phenomenology}
\label{sec:pheno}
%

\subsection{Enlarging the $M_N$ window}
In the standard type-I seesaw the parameters controlling leptogenesis are directly related to those entering the neutrino mass. As a result, the window for successful leptogenesis is restricted to be within $10^9 \lesssim M_N \lesssim 10^{15}$ GeV. The upper bound is obtained imposing $\Delta L = 2$ washout from scattering is small at temperatures of order $T\sim M_N$~(see for instance \cite{Buchmuller:2004nz}). The lower bound is derived requiring the CP violation parameter $\epsilon_N$ is large enough to reproduce the observed baryon asymmetry in the optimistic situation in which the efficiency factor is of order unity~\cite{Davidson:2002qv}.

In our hybrid model there is no strict connection between leptogenesis and neutrino masses (i.e., we have additional parameters), so can go beyond the aforementioned window. Combining the condition that off-shell scattering rate is slower than the Hubble rate [eq.~(\ref{scatt_wash})] with the neutrino mass formula [eq.~(\ref{mnu_IS})] we obtain:
\bea\label{upperB}
M_N \lesssim \left( \frac{16\pi^3 \sqrt{g_\ast} \; v^4}{M_{\rm Pl} \; m_\nu^2} \right)\times\left( \frac{y\langle \Phi_\lambda \rangle}{\kappa\langle \Phi_\kappa \rangle} \right)^4\sim10^{14}~{\rm GeV}\times\left( \frac{y\langle \Phi_\lambda \rangle}{\kappa\langle \Phi_\kappa \rangle} \right)^4.
\label{eq:upper_bound_M_N}
\eea
This bound is understood to constrain the largest combination of $\sim \lambda^2/M_N$ and does not depend on the assumption of anarchy of $\lambda$ or $M_N$ as we have done for eq.~\eqref{epsN}. The first factor in eq.~(\ref{eq:upper_bound_M_N}) is the result corresponding to the standard type-I seesaw. Careful numerical investigations show that scales as high as $M_N\sim10^{15}$ GeV are allowed \cite{Buchmuller:2004nz}. The second factor may be viewed as the result of a TeV-modulation and encapsulates the additional freedom our model features.  Nevertheless, in all these models $M_N$ is constrained to be smaller than the reheating temperature $T_{\rm reheat}$, otherwise it would not be produced efficiently. Current data suggests $T_{\rm reheat} \lesssim 10^{16}$ GeV \cite{Ade:2015tva}, so we will assume $M_N\lesssim 10^{16}$  GeV in the following.\footnote{The constraint is actually on the value of the Hubble scale at inflation, but may be translated into a bound on the reheating temperature assuming instantaneous reheating of a radiation dominated universe.} 

%
In order to derive the lower bound on $M_N$, we recall that it is conventional to write $Y_{\Delta B}\sim 10^{-3} \epsilon_N \eta$. In our model, the efficiency factor $\eta$ is a combination of washout from scattering and inverse decay in the UV [parametrized by $\Gamma_{\rm scattering}/H$ and $K_N$ of eq.~(\ref{KN}),
respectively] {\em and} washout at the TeV scale [with factor given by eq.~(\ref{IR_washout})]. Setting
$Y_{\Delta B} \sim 10^{-10}$ and $\eta \leq 1,$\footnote{One 
can have $\eta > 1$ only in the case of non-thermal production of $N$ where it can exceed its thermal abundance. However, here we will not consider this possibility.}
we get a familiar lower bound on the CP violation parameter, $\epsilon_N \gtrsim 10^{-7}$. Now, in our model $\epsilon_N$ is given by eq.~(\ref{epsN}). Plugging in the SM neutrino mass from eq.~(\ref{mnu_toy}) we then have
\bea
\epsilon_N \sim \frac{m_\nu M_N}{8 \pi v^2} \times \left( \frac{\kappa \left<\Phi_\kappa \right>}{y \left<\Phi_\lambda \right>} \right)^2 \gtrsim  10^{-7}.
\label{epsilon_model}
\eea
The CP asymmetry is thus the product of the same expression obtained in the standard seesaw and a TeV-modulation similar to the one appearing in eq.~(\ref{eq:upper_bound_M_N}). Importantly, the latter allows us to evade the Davidson-Ibarra bound \cite{Davidson:2002qv}, $M_N \gtrsim 10^9$ GeV, as may be seen by rewriting eq.~(\ref{epsilon_model}) as
\bea
M_N\gtrsim{10^{-7}}\frac{8\pi v^2}{m_\nu}
\left( \frac{y \left<\Phi_\lambda \right>}{\kappa \left<\Phi_\kappa \right>} \right)^2\sim{10^9~{\rm GeV}}
\times\left( \frac{y \left<\Phi_\lambda \right>}{\kappa \left<\Phi_\kappa \right>} \right)^2.
\label{M_N_lower_generic}
\eea
We emphasize that although eq.~(\ref{M_N_lower_generic}) is derived under the assumption of anarchical $\lambda$ and $M_N$, it holds even without such assumption and has therefore a general validity. There exists another rather generic lower bound $M_N \gtrsim 10^{6}$ GeV, which may be derived by requiring a sufficiently large CP violation and small enough $\Delta L = 2$ washout scattering \cite{Racker:2013lua}.

Assuming {\emph{anarchy}}, however, prevents us from going to such small values of $M_N$. Indeed, in that case we see that eqs.~(\ref{epsN}), (\ref{net_UV}), and (\ref{KN}) imply\footnote{Here we assume that washout due to scattering is under control, as in eq.~(\ref{scatt_wash}), although including such a suppression will not modify our argument below.}
\bea
~~~~~~~Y^N_{ \Delta \Psi }  \lesssim 10^{-3} \sqrt{ g_* } \frac{ M_N }{ M_{ \rm Pl } }~~~~~~~~~({\rm anarchic~regime}).
\label{Y_UV_upper}
\eea
Since our model has additional washout at the TeV scale [second factor on RHS of eq.~(\ref{final-asym})], we need $Y^N_{ \Delta \Psi } \gtrsim 10^{-10}$. Combining the latter with eq.~(\ref{Y_UV_upper}) we get
\bea
~~~~~~~~~M_N \gtrsim 10^{ 11 } \hbox{GeV}~~~~~~~~~~~~~~~({\rm anarchic~regime}).
\label{M_N_lower_stronger}
\eea
Eq.~(\ref{M_N_lower_stronger}) is stronger than eq.~(\ref{M_N_lower_generic}) because of the tight relation
between $K_N$ in eq.~(\ref{KN}) -- which enters $Y^N_{ \Delta \Psi }$ via $\eta$ -- and $\epsilon_N$ in eq.~(\ref{epsN}).
The only way to evade the bound in eq.~(\ref{M_N_lower_stronger})  is by 
relaxing the assumptions made in obtaining it, for instance by allowing some hierarchies in $\lambda$'s and/or $M_N$'s of different $N$ generations
so that we have more freedom to adjust $\eta$ and $\epsilon_N$ independently.\footnote{Alternatively, as in the case of the Davidson-Ibarra bound in standard seesaw, one can relax the lower bound by having quasi-degenerate right-handed neutrino mass spectrum (see for instance \cite{Deppisch:2010fr}) and/or taking into account of lepton flavor effects~\cite{Racker:2012vw}. While the former calls for new ingredient to explain the quasi-degeneracy, the latter requires the neutrino Yukawa couplings to be very hierarchical among different $\Psi,\Psi^c$ flavors.} In particular, allowing hierarchies of order $10^{-4}-10^{-3}$ in the couplings one can show that values as low as $M_N \gtrsim 10^{6}$ GeV, one of the generic lower bounds mentioned above, are possible in our model~\cite{long}. Note that such a light $N$ is especially welcome in local supersymmetric theories in order to avoid the gravitino problem~\cite{Pagels:1981ke}; numerical studies in this case impose $M_N\lesssim10^9$ GeV for either stable~\cite{Moroi:1993mb} or unstable gravitino~\cite{Kawasaki:2008qe}.


\begin{figure}[t]
\centering
\includegraphics[width=75mm,height=80mm]{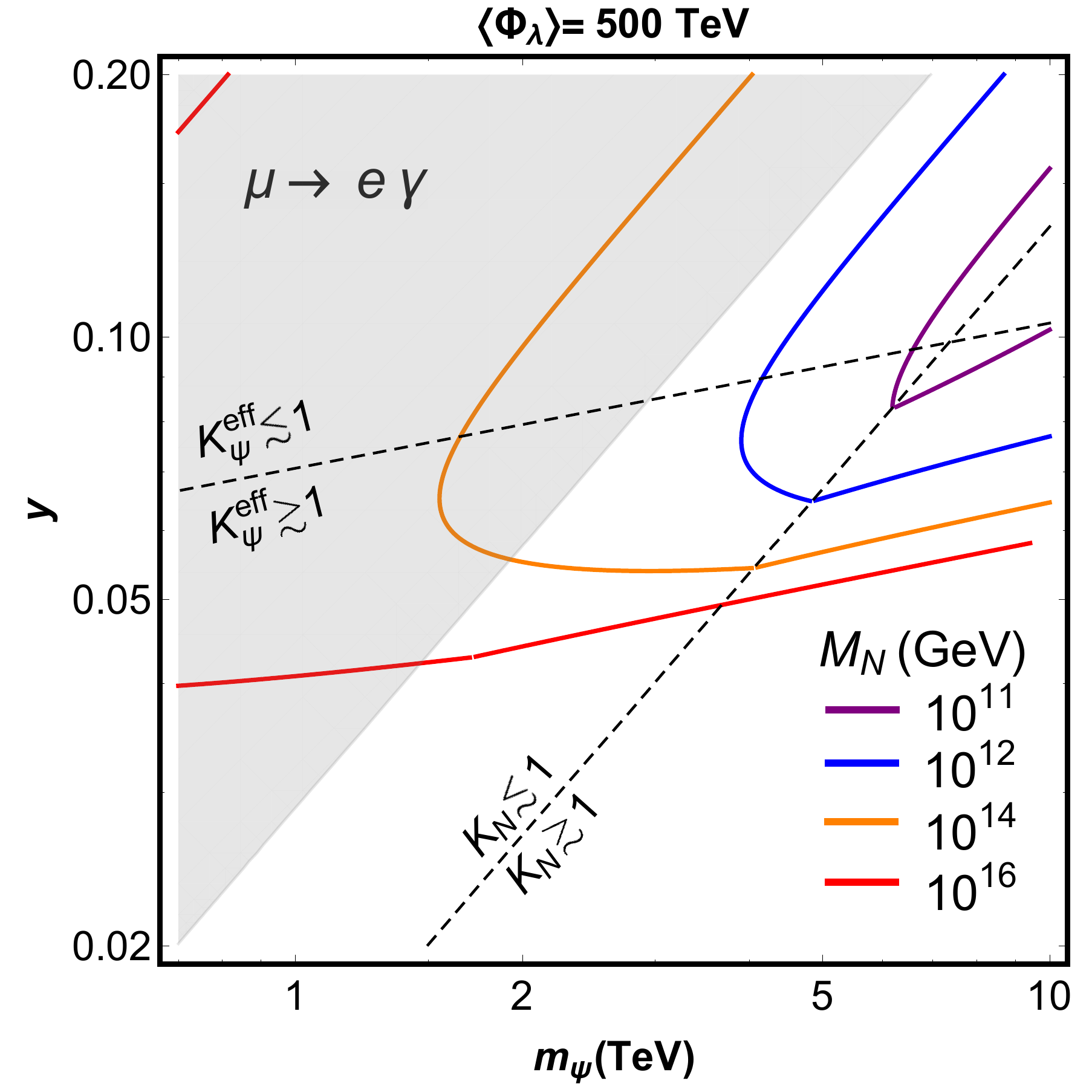}  \includegraphics[width=75mm,height=80mm]{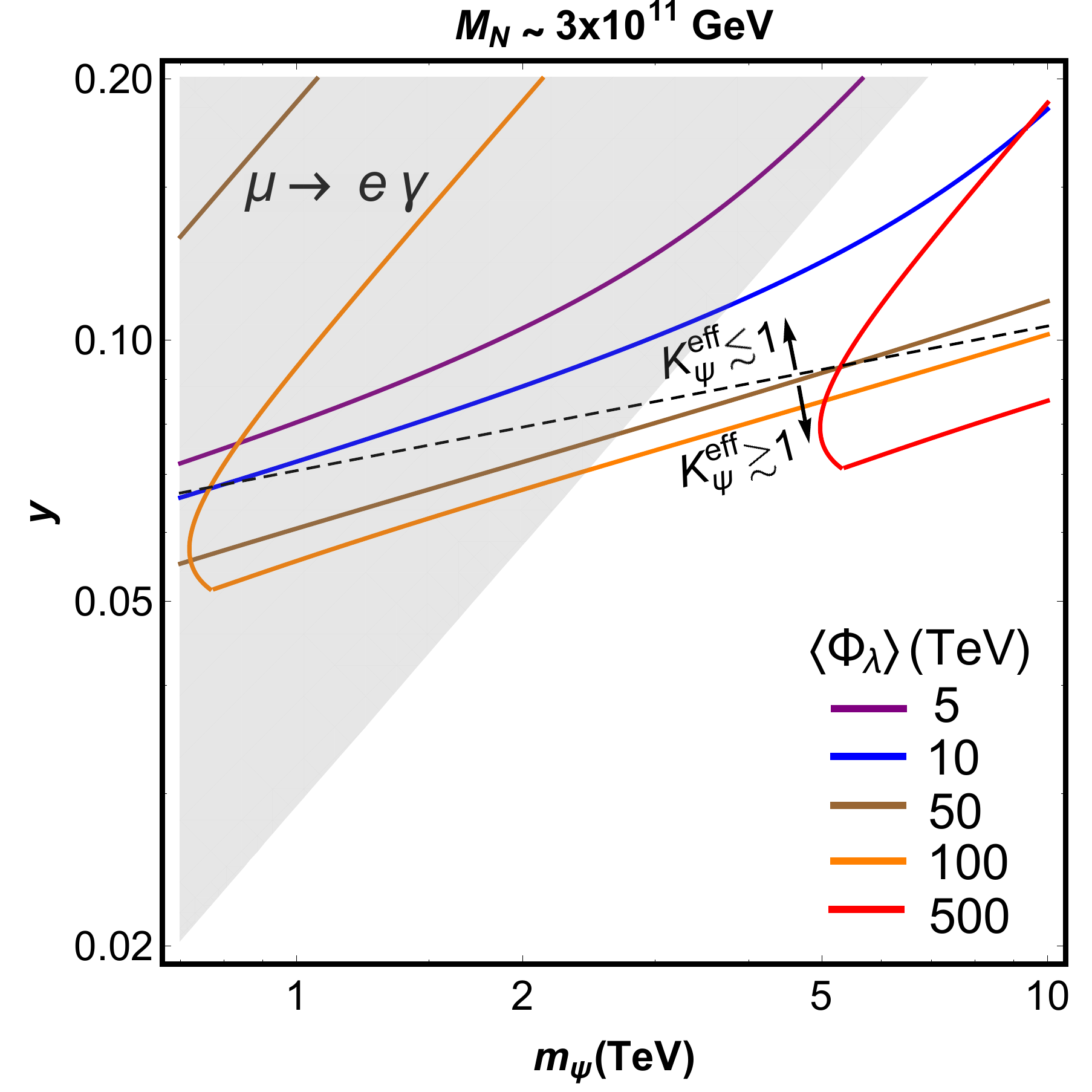}
\caption{For each solid curve 
in the plane of Yukawa coupling ($y$) and mass ($m_{ \Psi }$) of the pesudo-Dirac singlet fermion, 
the observed baryon asymmetry and neutrino masses are reproduced. Couplings and mass matrices are assumed to be non-hierarchical; specifically $M_{N_{1,2}}$ are taken to be of the same order of magnitude but non-degenerate. In the left panel we present different choices of $M_N$ with fixed $\langle\Phi_{ \lambda }\rangle = 500$ TeV. In the right panel we vary $\langle\Phi_{ \lambda }\rangle$ and keep $M_N \sim 3 \times 10^{11}$ GeV fixed. The dashed lines set the boundary between the weak and strong washout regimes in the UV or IR, parametrized by $K_N$ or $K_\Psi^{\rm eff}$ respectively. The kinks on the curves are the artifacts of our interpolation between the strong and weak washout regimes. The gray shaded region is excluded by the bound from $\mu\to e\gamma$.}
\label{fig:Plot1}
\end{figure}

A quantitative analysis of the parameter space compatible with a successful leptogenesis is presented in the $m_\Psi-y$ plane for different choices of $M_N$ and $\langle\Phi_{ \lambda }\rangle$ in figure~\ref{fig:Plot1}. They are produced, under the assumptions made below eqs.~(\ref{epsN}) and (\ref{net_UV}), using more accurate formulae derived in \cite{long}.
On each curve the observed final asymmetry and neutrino mass are obtained. The plot on the left panel shows curves for a fixed  $\langle\Phi_{ \lambda }\rangle$ and different values of $M_N$. The plot on the right panel shows curves for a fixed $M_N$ and different values of $\langle\Phi_{ \lambda }\rangle$. Overall we see that we can obtain the observed baryon asymmetry and neutrino masses over a wide range of parameters.

As an illustration, let us discuss the case $M_N \sim 10^{16}$ GeV (see the solid red curve in the left panel of figure \ref{fig:Plot1}) and leave a detailed exploration of low $M_N$ scenario for \cite{long}. For instance, we can choose $0.4 \lesssim \lambda \lesssim 1$ such that $1 \lesssim K_N \lesssim 9$ [see eq.~\eqref{KN}], which satisfies the condition of washout from scattering being under control, i.e., eq.~(\ref{scatt_wash}).
Then, the UV asymmetry [see eq.~\eqref{net_UV}, combined with eqs.~(\ref{epsN}) and (\ref{KN})] 
is given by $\sim 10^{ -3 } \sqrt{ g_* } M_N / M_{ \rm Pl }$ so that it too large by several orders of magnitude
for $M_N \sim 10^{ 16 }$ GeV. 
Yet, our model can generate a realistic present-day asymmetry taking advantage of a strong IR washout. This may be achieved with $y\lesssim 0.05$ (note the strong sensitivity of eq.~(\ref{IR}) on $y$). This possibility should {\em not} be viewed as fine-tuning. Indeed, the baryon asymmetry [eq.~(\ref{final})] is the product of two numbers that are typically smaller than 1, but have no favored value a priori. The observed $Y^{\rm obs}_{\Delta B}\sim10^{-10}$ (that is {\emph{not}} a special number in any sense) can naturally be obtained suppressing one or both factors in eq.~(\ref{final}): an interplay between UV and IR effects is a generic feature of our model.
The SM neutrino mass can also be reproduced with $M_N \sim 10^{16}$ GeV. Indeed, even though a large $M_N$, $\lambda \lesssim1$, and $y \lesssim 0.05$ tend to suppress eq.~(\ref{mnu_toy}), we still have the freedom to assume $\langle\Phi_{ \lambda }\rangle > 1$ TeV. 
Importantly, the latter parameter {\emph{does not}} control any particle mass in the model. In particular, 
even with a rather large $\langle\Phi_{ \lambda } \rangle$,
the singlets $\Psi,\Psi^c$ can still be at the TeV scale as long as $\kappa\langle\Phi_{ \kappa }\rangle, (\lambda\langle\Phi_{ \lambda }\rangle)^2/M_N\lesssim1$ TeV. As stressed below eq.~(\ref{mnu_toy}), the only physical effect of $\langle\Phi_{ \lambda }\rangle/\langle\Phi_{ \kappa }\rangle$ is to modify the SM neutrino mass formula compared to the type-I. 
In the warped/composite model, an ``effective'' modulation factor (much) larger than $O(1)$ can be readily obtained, say with natural size of the fundamental parameters, and without requiring a commensurate hierarchy of mass scales.

\subsection{Constraints and signatures}

Let us now consider the most important, and generic constraints and signals on our hybrid seesaw model.

First of all, the rare process $\mu \to e\gamma$ is severely constrained. Assuming anarchic Yukawa couplings $y$ and $m_\Psi$, the branching ratio of $\mu \to e\gamma$ can be written as \cite{Cheng:1980tp}
\bea
\textrm{BR}(\mu \to e \gamma)\simeq \frac{3\alpha_{\rm em}}{8\pi}\left(\frac{y v}{m_\Psi}\right)^4,
\eea
where $\alpha_{\rm em}\approx 1/137$ is the fine structure constant, $v\approx174$ GeV, and we neglected terms of order $m_W^2/m_\Psi^2$. The current experimental bound $\textrm{BR}(\mu \to e \gamma)< 4\times 10^{-13}$ \cite{TheMEG:2016wtm} translates into $y /m_{\Psi} \lesssim 2.7 \times 10^{-2} / {\rm TeV}$ (see the gray shaded region in figure~\ref{fig:Plot1}). Allowing a mild hierarchy in $y$ relaxes the bound further. Other constraints on $y,m_\Psi$ are much weaker and will not be considered. 

The Higgs portal couplings $g^2_{H\Phi_\kappa, H\Phi_\lambda}|H|^2|\Phi_{\kappa,\lambda}|^2$ lead to a mixing angle between the SM Higgs and the new scalars of order $g^2_{H\Phi}v\langle\Phi\rangle/m^2_\Phi$. Requiring this is below $\sim10\%$ \cite{Robens:2016xkb}, and assuming couplings of order unity, we find our model is consistent with data if the new scalar masses are in the TeV range.

The collider signatures of our TeV singlet fermions are similar to those of standard inverse seesaw models, where $\Psi,\Psi^c$ are produced in association with SM leptons via off-shell $W$. Final states will be opposite sign dileptons, accompanied by jets or trileptons, with missing transverse momentum (see for example \cite{Deppisch:2015qwa} and references therein). The heavy scalars $\Phi_{\kappa,\lambda}$ might be produced via the Higgs portal, and decay into SM via the Higgs or into $\Psi,\Psi^c$ pairs, if kinematically allowed.

\subsubsection{Constraints and signatures on fully-symmetric models}

In the fully-symmetric models there are other constraints to be taken into account. If none of the symmetries in table~\ref{tab} are gauged, these scenarios predict two massless NGBs. One of the two is especially problematic because its couplings to the SM are relatively unsuppressed. Fortunately, these scenarios emerge from a UV completion in which the associated, i.e., $U(1)_{B-L}$, symmetry is gauged~\cite{long}: this NGB is therefore unphysical. 
We then need to consider the phenomenology of the heavy $B-L$ vectors. The main constraint arises from precision electroweak bounds. To avoid any tension with data we may assume $\langle\Phi_\kappa\rangle\gtrsim7$ TeV~\cite{Cacciapaglia:2006pk}. With the choice $\kappa\sim0.1$, our singlets $\Psi,\Psi^c$ have masses in the TeV range, as desired. The $U(1)_{B-L}$ gauge boson could also lead to interesting signals at colliders, being produced via quark fusion, and then decaying into jets and TeV singlet fermions pairs if kinematically allowed \cite{Deppisch:2015qwa}.

The remaining NGB, $\pi_\lambda$, is physical and emerges from the spontaneous breaking of the $U(1)_\lambda$ symmetry in table~\ref{tab}. The crucial point however is that $\pi_\lambda$ has a very weak coupling to the SM. This can be obtained by transforming all fermions via a {\emph{local}} $U(1)_\lambda$ rotation by an angle $\pi_\lambda/\langle\Phi_\lambda\rangle$. This procedure removes all non-derivative couplings of the NGB but introduces a coupling of the fermion currents to $\partial_\mu\pi_\lambda$. After having integrated out the heavy fields we are left with the SM Lagrangian plus
\bea
\delta{\cal L}_{\rm EFT}&=&\frac{1}{2}(\partial_\mu\pi_\lambda)^2-\frac{\pi_\lambda}{\langle\Phi_\lambda\rangle} \partial_\mu J_{B-L, {\rm SM}}^\mu+\cdots, 
\eea
where the dots refer to higher dimensional interactions involving the SM fields and derivatives of $\pi_\lambda$. The integration by parts converts the coupling of the NGB into an interaction with the higher-dimensional operators that violate the SM $B-L$. The leading one is the neutrino mass operator, giving a coupling $y_\pi\nu\nu\pi_\lambda$, where 
\bea\label{ypi}
y_\pi=\frac{m_\nu}{\langle\Phi_\lambda\rangle}\sim10^{-13}\,\frac{{\rm TeV}}{\langle\Phi_\lambda\rangle}.
\eea
With such a tiny coupling, $\pi_\lambda$ is expected to be consistent with astrophysics bounds even if exactly massless \cite{Gelmini:1982rr}. Furthermore, $\pi_\lambda$ typically decouples before the QCD phase transition, and it gives negligible corrections to $\Delta N_{\rm eff}$ \cite{Ade:2015xua}, see e.g. \cite{Brust:2013xpv}. Our NGB has many properties in common with the Majoron studied in~\cite{Gelmini:1980re}. 

Higher dimensional operators can explicitly break $U(1)_\lambda$ and contribute a small mass for $\pi_\lambda$. Depending on the mass there could be a variety of cosmological and astrophysical signatures, see e.g. ref.~\cite{Chacko:2003dt}. Finally, the SM Higgs could decay to a pair of $\pi_\lambda$, contributing to invisible decays of the SM Higgs \cite{Joshipura:1992ua}. 
This is consistent with the current bounds on this process for $\langle\Phi_\lambda\rangle\gtrsim1$ TeV \cite{Bonilla:2015uwa},
as in our benchmarks, while allowing future detection.

\section{Outlook}

Combining the inverse seesaw with a high-scale seesaw module, one can \emph{naturally} explain light neutrino masses and obtain a successful leptogenesis within a testable framework. This is achieved via a {\emph{non-generic structure}} involving TeV-scale particles acting as mediators between the super-heavy Majorana singlet fermion and the SM. We identified the ingredients necessary to obtain viable models and studied a class of simple realizations. The absence of direct couplings between the super-heavy singlet and the SM, that is crucial in our picture, can be enforced by a gauge $U(1)_{B-L}\times U(1)_{X}$ symmetry~\cite{long} or naturally emerge from the framework of a warped extra dimension/composite Higgs, as shown by some of us in a previous paper \cite{Agashe:2015izu}. 

We find that these scenarios have a very rich phenomenology. Relaxing the connection between neutrino masses and $M_N$, see eq.~(\ref{mnu_toy}), allows us to enlarge the parameter space consistent with leptogenesis. Singlets heavier than $\sim 10^{15}$ GeV or lighter than $\sim 10^9$ GeV become possible while still accommodating realistic neutrino masses: we explicitly showed the former case in this paper, while for the latter situation we do need a hierarchy between 1st and 2nd generation singlet Yukawa couplings, as we will see in \cite{long}. Furthermore, the presence of a high scale and low scale module implies that these constructions are characterized by four different washout regimes: strong-UV/strong-IR, strong-UV/weak-IR, weak-UV/strong-IR and weak-UV/weak-IR. An interplay between UV genesis and IR physics is therefore a very distinctive property of these models, as seen in figure \ref{fig:Plot1}. Overall, the above features characterize a new paradigm for leptogenesis.

Finally, our scenarios have interesting experimental signatures besides those induced by the TeV singlet fermions of the more conventional inverse seesaw. Indeed, all our models have new {\em scalars} within the reach of present and future colliders. Also, certain realizations predict a light (pesudo-)scalar coupled dominantly to neutrinos, which may have interesting cosmological effects, as well as a new vector boson associated to the gauged $U(1)_{B-L}$ symmetry --- introduced to obtain our model eq.~(\ref{model}) within a weakly-coupled 4D theory. A more detailed analysis of these scenarios will be presented in~\cite{long}.

\section*{Acknowledgements}

We would like to thank Csaba Csaki, Bhupal Dev, Andre de Gouvea, Pedro Machado, Rabindra Mohapatra, Matthias Neubert, Qaisar Shafi, Geraldine Servant, Jessica Turner and Yue Zhang for discussions.
The work of KA, PD, ME and SH was supported in part by NSF Grant No.~PHY-1620074 and the Maryland Center for Fundamental Physics. 
KA and PD were 
also supported by the Fermilab Distinguished Scholars Program and 
SH by NSF Grant No.~PHY-1719877. 
CSF was supported by the S\~{a}o Paulo Research Foundation (FAPESP) grants 2012/10995-7 \& 2013/13689-7 and 
is currently supported by the Brazilian National Council for Scientific and Technological Development (CNPq) grant 420612/2017-3.
He also acknowledges the hospitality of Fermilab theory group during his visit supported by FAPESP grant 2017/02747-7 when part of the work was carried out.
LV is supported by the Swiss National Science Foundation under the Sinergia network CRSII2-16081.
%


 \end{document}